\documentclass[showpacs,aps,floatfix,prb,reprint,superscriptaddress]{revtex4-1}
\usepackage{xfrac}
\usepackage{amssymb}
\usepackage{braket}
\usepackage{graphicx}
\usepackage{verbatim}
\usepackage{etoolbox}
\usepackage{amsfonts} 
\usepackage{amsmath} 
\usepackage[utf8]{inputenc} 
\usepackage{mathtools}
\usepackage{soul,xcolor,colortbl}
\usepackage{hyperref} \hypersetup{colorlinks=true,citecolor=blue,linkcolor=blue,urlcolor=black} 
\usepackage{morefloats}
\usepackage{natmove}  

\usepackage{bm} 
\usepackage{array}
\newcolumntype{C}[1]{>{\centering\arraybackslash}p{#1}}\usepackage{soul}
\usepackage{color} 
\def\AFLOW{{\small AFLOW}}

\def\TI{{\small TI}}
\def\TIs{{\small TI}s}
\def\SI{Supplementary Information}
\def\SOC{{\small SOC}}
\definecolor{pranab_green}{rgb}{0.31,0.53,0.10}
\definecolor{pranab_red}{rgb}{0.85,0.23,0.11}
\definecolor{PaleBlue}{rgb}{0.4,0.7,1}

\def\citeAFLOWALL{\cite{curtarolo:art65,curtarolo:art110,curtarolo:art104, curtarolo:art63,curtarolo:art57,curtarolo:art58,curtarolo:art49,curtarolo:art87,curtarolo:art127,curtarolo:art121,curtarolo:art75,curtarolo:art92,curtarolo:art128}}

\makeatletter \renewcommand\frontmatter@abstractwidth{\dimexpr\textwidth\relax} \makeatother  

\begin{document}
\title{\Large Spinodal superlattices of topological insulators}

\author{Demet Usanmaz}
\affiliation{Department of Mech. Engineering and Materials Science, Duke University, Durham, NC 27708, USA.}
\author{Pinku Nath}
\affiliation{Department of Mech. Engineering and Materials Science, Duke University, Durham, NC 27708, USA.}
\author{Cormac Toher}
\affiliation{Department of Mech. Engineering and Materials Science, Duke University, Durham, NC 27708, USA.}
\author{Jose Javier Plata}
\affiliation{Department of Mech. Engineering and Materials Science, Duke University, Durham, NC 27708, USA.}
\author{\\Rico Friedrich}
\affiliation{Department of Mech. Engineering and Materials Science, Duke University, Durham, NC 27708, USA.}
\author {Marco Fornari}
\affiliation{Department of Physics, Central Michigan University, Mount Pleasant, MI 48859, USA.}
\author {Marco Buongiorno Nardelli}
\affiliation{Department of Physics, University of North Texas, Denton, TX 76203, USA. }
\author{Stefano Curtarolo}
\email{stefano@duke.edu}
\affiliation{Materials Science, Electrical Engineering, Physics and Chemistry, Duke University, Durham, NC 27708, USA.}
\affiliation{Fritz-Haber-Institut der Max-Planck-Gesellschaft, 14195 Berlin-Dahlem, Germany.}

\date{\today}

\begin{abstract}
\noindent
Spinodal decomposition is proposed for stabilizing self-assembled interfaces between topological insulators (TIs)
by combining layers of iso-structural and iso-valent TlBi$X_2$ ($X$=S, Se, Te) materials.
The composition range for gapless states is addressed concurrently to
the study of thermodynamically driven boundaries.
{
  By tailoring composition, the TlBiS$_2$-TlBiTe$_2$ system might produce both spinodal
  superlattices and two dimensional eutectic microstructures,
  either concurrently or separately.
}
The dimensions and topological nature of the metallic channels are
determined
by following the spatial distribution of the charge density and the spin-texture.
The results validate the proof of concept for obtaining {\it
  spontaneously forming} two-dimensional TI-conducting channels
embedded into three-dimensional insulating environments without any
vacuum interfaces.
Since spinodal decomposition is a controllable kinetic phenomenon, its
leverage could become the 
long-sought enabler for effective TI technological deployment.
\end{abstract}

\maketitle

\section*{Introduction}

Modern materials technology has been efficiently used to reduce the size and increase the functionality of electronic devices and electrical machinery. 
Semiconductors find a place in a wide range of applications, such as transistors~\cite{Cressler_2003}, detectors~\cite{Razeghi_jap_1996,Rizzi_jas_2010}, light emitting diodes~\cite{Hwang_apl_2005}, and lasers~\cite{Kohler_nature_2002}; while powerful permanent magnets are used in renewable power generation and electric motors central to the post fossil fuel economy~\cite{curtarolo:art81}.
Many of these technologies require scarce elements to achieve exotic properties, such as the reliance of transparent conductors on indium~\cite{exarhos_discovery-based_2007}, or the dependence of permanent magnets on rare earth elements, {\it e.g.} dysprosium~\cite{curtarolo:art97,Nakamura_NMAT_materials_scarcity_2011}. 
Similar properties can be replicated by fabrication of heterostructures, where the distinct band structures of the constituent materials facilitates the engineering of electronic properties inside device components.
However, this technology relies on artificial growth methods, which increase both the time and cost of fabrication. 
Exploring alternative synthesis methods and materials is important for overcoming this challenge~\cite{curtarolo:art107}. 

Here, we propose a new approach by combining advanced thermodynamic and electronic structure concepts for novel materials design.
The design and manufacture of embedded 2D metallic channels in a 3D insulating matrix will enhance the functionality of scalable circuitry for high-performance electronics.
Our approach to obtain these channels exploits easy-to-make self-organizing thermodynamically driven morphologic microstructures.
This is an alternative to current efforts, where fabrication of heterostructures has relied mostly on artificial growth methods, very likely with thermodynamically unstable or metastable phases~\cite{Hirahara_prl_2011,Berntsen_prb_1013,Shoman_nc_2015}.
Topological insulators (\TIs),
which exhibit insulating behavior in bulk and metallic states at their boundaries, are the most promising materials for this purpose~\cite{Yan_rpp_2012,Chen_science_2009,Zhang_Nat.Phys._2009,zhang_PRL_2009,curtarolo:art77}.
The presence of metallic states relies on the spin-orbit induced band inversion in bulk materials, protected by either time-reversal or crystal symmetry~\cite{zhang_PRL_2009,Hsieh_natcom_2012,Dziawa_NatMAt_2012,Tanaka_NatPhy_2012,Ando_jpsj_2013}.
The characteristics of the gapless boundary states are linear dispersion in the bulk band gap, spin-texture, robustness against scattering by non-magnetic impurities, and symmetry protection.
Studies have demonstrated that the formation of heterostructures~\cite{Nakayama_PRL_2012,Rauch_prb_2013,Wu_scirep_2013,Chen_nanolett_2015}, alloying~\cite{Xu_science_2011,Dziawa_NatMAt_2012}, and thickness engineering~\cite{Zhang10natphys} have advantages for controlling the electronic properties of \TIs.
In addition, recent studies show that it is possible to observe novel properties in \TI\ superlattices, such as both time-reversal and crystal symmetry protected surface states~\cite{Eschbach_nc_2017,Weber_prl_2015}, band structure tuning through a topological phase transition~\cite{Belopolski_sciadv_2017}, 
{
topologically nontrivial surface states in a magnetic-TI/TI superlattice~\cite{Hagmann_njp_2017},
}
and the realization of 3D Weyl semimetal phases~\cite{Burkov_prl_2011}.
To fully exploit \TIs\ in future devices, a detailed exploration of \TI\ heterostructures/superlattices is needed.

The natural formation of interfaces based {on} thermodynamic stability is interesting:
recent studies have proposed the use of spinodal decomposition to generate materials with {\it ad hoc} characteristics which enhance the properties of thermoelectric materials~\cite{Androulakis_jacs_2007,Zhao_jacs_2012,Gelbstein_CHEMMAT_Spinodal_2010,Sootsman_JAP_2009}.
Spinodal decomposition is a phase separation mechanism, where below the critical temperature $T_{\mathrm c}$, the components separate into distinct homogeneous regions with different physical and chemical properties~\cite{Hillert1961525,Cahn1961795,Cahn1962179}.
This allows \TI\ heterostructures consisting of layers of thermodynamically stable phases to be constructed, which is the key to durable and long-lasting applications.
Moreover, creating boundaries between chemically distinct but iso-structural phases of some insulating materials has already led to remarkable properties~\cite{mannhart_oxide_2010,yoshimatsu_dimensional-crossover-driven_2010,yoshimatsu_origin_2008}.

The candidate materials, (Tl,\hspace{0.1cm}Bi)-based ternary chalcogenides, TlBi{\it X}$_2$ ({\it X}=S, Se, Te), are a rare group in which \TIs\ and trivial-insulators share the same crystal structure.
TlBiSe$_2$ and TlBiTe$_2$ are 3D \TI\ materials while TlBiS$_2$ is a trivial insulator. 
These properties allow the investigation of the \TI/\TI~ and trivial-insulator/\TI ~boundaries between phases that are both iso-structural (important for forming commensurate interfaces) and iso-valent.
Furthermore, TlBiSe$_2$ is a \TI\ similar to Bi$_2$Se$_3$ with the advantage of having an almost perfect Dirac cone at the $\Gamma$-point, which is separated from the bulk states and thus allows for the investigation of the transport properties of the TlBiSe$_2$ boundary states independently of the bulk states~\cite{Analytis_prb_2010, Eto_prb_2010, Butch_prb_2010}.
The topologically trivial system, TlBiS$_2$, has a clear band gap at the surface.
Gapless metallic states arise at the onset of the topological phase transition, occurring under strain and/or an external electric field ~\cite{Yan_epl_2010,Zhang_sr_2015,Singh_jap_2014}.
Another interesting characteristic of this group of materials is the topological phase transition from trivial-insulator to \TI\ resulting from substituting S with Se atoms in TlBi(S$_{1-x}$Se$_x$)$_2$~\cite{Xu_science_2011}.
This tunes the lattice constant and spin-orbit coupling (\SOC), so that a topological phase transition takes place at the critical {point} ($x_{\mathrm c}$ $\sim$ 0.48) and a 3D-Dirac point arises.
{For} ($x_{\mathrm c}$ $\le$ $x$ $\le$ 1), gapless surface states appear in the electronic band gap~\cite{Xu_science_2011}.
In contrast, other independent studies report that the TlBi(S$_{1-x}$Se$_x$)$_2$ system has gapped surface states for 0.6 $\le$ $x$ $\le$ 0.9~\cite{Sato_np_2011,Souma_prl_2012}, despite being in the topologically non-trivial phase.
The physical cause of the opening of the gap is not known, although it is possible that the answer lies in the thermodynamic properties.
Additionally, Weyl semimetal phases are predicted for TlBi(S$_{1-x}$Se$_x$)$_2$ and TlBi(S$_{1-x}$Te$_x$)$_2$ alloys at $x=0.5$ in the case of layer-by-layer growth (Tl-Se(Te)-Bi-S) for certain critical values of the $c/a$ ratio~\cite{Singh_prb_2012}.
Other interesting phenomena, specifically Rashba spin-splitting and the topological proximity effect, have been observed on the Bi(1 bilayer)/TlBiS$_2$ and Bi(1 bilayer)/TlBiSe$_2$ surfaces, respectively~\cite{Shoman_nc_2015}.
These interesting properties make this group of materials promising candidates for superlattice studies.

In this work, we combine thermodynamic and electronic structure analysis to define a natural design strategy for \TI\ superlattices that can be an alternative to the very costly and time-consuming experimental artificial growth methods.
The current study is built on the following aspects:
{\bf i.} Identify the miscibility gap of each pair of TlBi$X_2$ ($X$=S, Se, Te) compounds by calculating the thermodynamic phase diagram and predicting the consolute temperature ($T_{\mathrm c}$).
{\bf ii.} Design boundaries between constituent materials to predict appropriate ranges of composition guaranteeing the topologically protected gapless metallic states.
{\bf iii.} Track the spatial distribution of the charge density of these boundaries to verify the existence of topologically protected 2D metallic channels in the 3D insulating matrix.

\section*{\label{sec:method}Methods}

\noindent
{\bf Thermodynamics.}
The enumeration of configurations, determination of ground state structures, and prediction of energies of more structures through Cluster Expansion 
are performed within the {\small ATAT}
framework~\cite{Walle_calphad_2002,atat1,axel_MC}.
The temperature-composition phase boundaries are obtained with Monte Carlo simulations~\cite{Walle_calphad_2002}.
From this data, using the definition of the Gibbs free energy in Ref.~\cite{curtarolo:art107}, proper derivatives are calculated to determine the binodal and spinodal {\it loci}.

\noindent
{\bf Bulk.}
Bulk structures are fully relaxed using the \AFLOW\ high-throughput framework~\citeAFLOWALL\ and the {\small DFT} Vienna {\it Ab-initio} Simulation Package ({\small VASP})~\cite{kresse_vasp}.
Geometry optimizations are performed following the \AFLOW\ standard~\cite{curtarolo:art104}, using {\small PAW} pseudopotentials~\cite{PAW} and the {\small PBE} parametrization of the {\small GGA} exchange and correlation functional~\cite{PBE}.
A high energy-cutoff (40$\%$ larger than the maximum cutoff of all pseudopotentials) is used for all calculations.
Reciprocal space integration is performed using a mesh of 8000 {\bf k}-points per reciprocal atom.
Structures are fully relaxed (cell volume and ionic positions) until the energy difference between two consecutive ionic steps is smaller than $10^{-4}$ eV. 
Electronic structure calculations for TlBiS$_2$, TlBiSe$_2$, and TlBiTe$_2$ are performed with \SOC.

{\noindent
\bf Interfaces.}
The lattice parameters (stress) and ionic positions (forces) are optimized until all force components on each ion are less than 0.001 eV/{\AA}. 
The outer $s$ and $p$ electrons are treated as valence electrons with the rest included in the core.
A kinetic energy cutoff of 392 eV and a $\Gamma$-centered $9 \times 9 \times 1$ {\bf k}-point mesh is used.
Electronic structure calculations are performed with \SOC.

\section*{\label{sec:results}Results and Discussion}

\begin{table}[h]
\caption{\small
 Cross-validation ({\small CV}) score (in meV), consolute temperatures ($T_{\mathrm c}$) (in K), and critical compositions ($x_{\mathrm c}$) of TlBiS$_2$-TlBiSe$_2$, TlBiSe$_2$-TlBiTe$_2$, and TlBiS$_2$-TlBiTe$_2$.}
\centering
\begin{tabular}{ l | c c c }
 \hline			
 ~ & {\small TlBiS$_2$-TlBiSe$_2$} & {\small TlBiSe$_2$-TlBiTe$_2$} & {\small TlBiS$_2$-TlBiTe$_2$} \\
 \hline			
 {\small CV} score & 0.63 & 1.15 & 4.30 \\
 $T_{\mathrm c}$ & 162 & 400 & 1040 \\
 $x_{\mathrm c}$ & 0.4 & 0.4 & 0.35 \\
 \hline			
\end{tabular}
\label{Table-CV}
\end{table}
\noindent

\begin{figure*}[]%
  \includegraphics*[width=0.99\textwidth,clip=true]{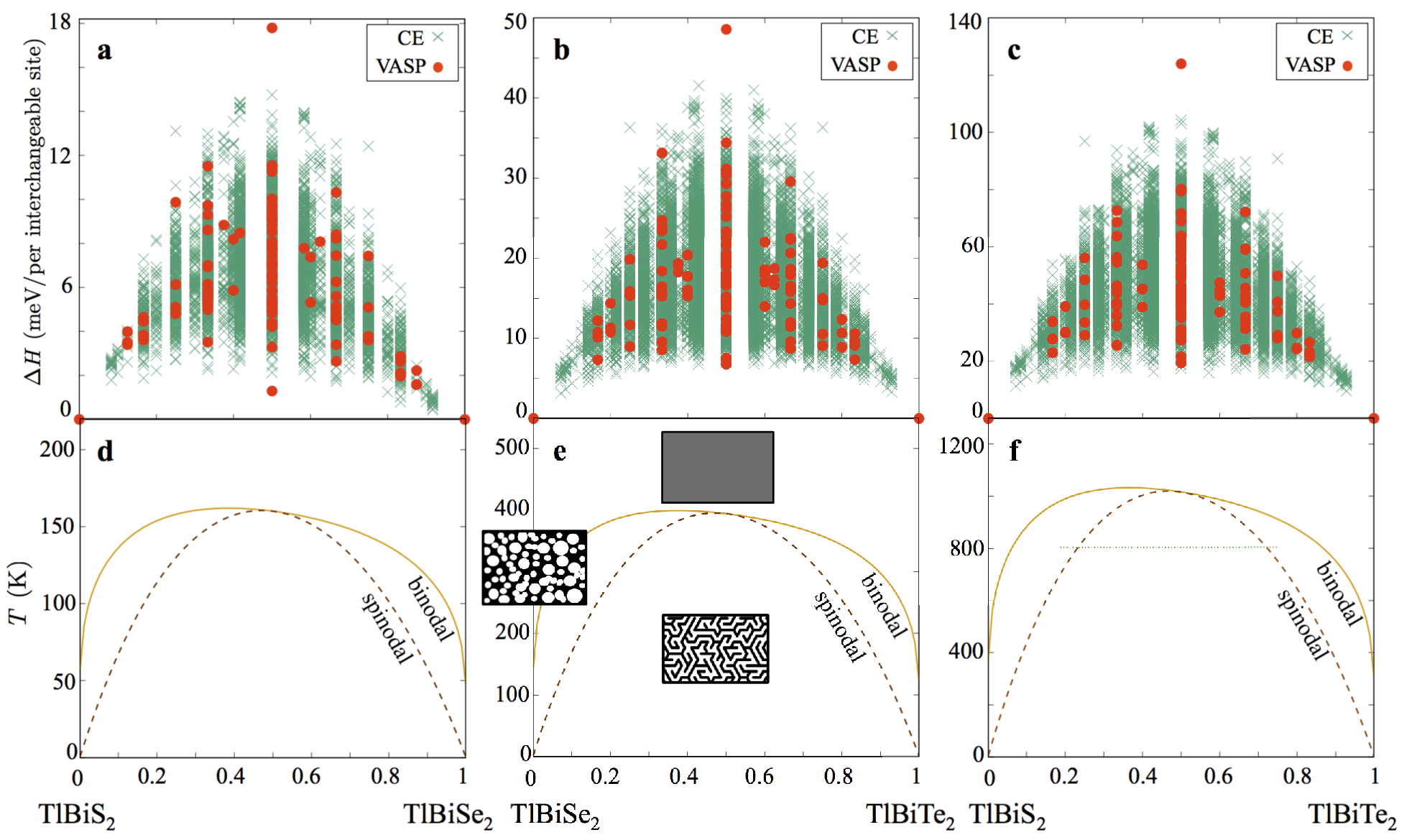}
  \vspace{-2mm}
  \caption{\small
    Formation enthalpies of the
    {\bf (a)} TlBiS$_2$-TlBiSe$_2$ 
    {\bf (b)} TlBiSe$_2$-TlBiTe$_2$ 
    {\bf (c)}TlBiS$_2$-TlBiTe$_2$ structures using DFT calculations (\textcolor{pranab_red}{$\bullet$}) and cluster expansion technique (\textcolor{pranab_green}{$\times$}) (upper panel).
    Their respective binodal ({\color{orange}{---}}) and spinodal curves ({\color{brown}{- - -}}) are illustrated in the lower panels {\bf (d,e,f)}.
    {The eutectic isotherm, from Ref.~\cite{Jafarov_im_2014,Babanly_amchemscij_2016} is illustrated with} ({\color{pranab_green}{$\cdot\cdot\cdot$}}).
  }
  \label{fig1}
\end{figure*}

{\bf Thermodynamic properties.}
Figure~\ref{fig1}(a, b, c) illustrates the calculated and predicted formation enthalpies ($\Delta H_{\mathrm f}$) from {\small DFT} and cluster expansion, respectively.
All three systems have positive $\Delta H_{\mathrm f}$ for the entire range of compositions ($0\!\! <\!\!x\!\! <\!\!1$): these systems are immiscible at 0K.
Cluster expansion predictions well agree with {\small DFT} calculated energies with small cross validation scores which are tabulated in Table~\ref{Table-CV}.
The miscibility of two isomorphous systems can be addressed using the classic Hume-Rothery rules~\cite{cottrell_1967} based on four properties: atomic radius, crystal lattice, valence, and electronegativity.
These three systems have the same crystal structure; S, Se, and Te atoms are from the same group in the periodic table and have similar electronegativities; while the atomic radii range from 1.04 {\AA} for S, to 1.17 {\AA} for Se, and 1.37 {\AA} for Te.
This difference creates a mismatch between the constituent lattices, reducing interface coherence, eventually leading to phase decomposition.

Combining the output of the Monte Carlo simulations with a previously developed model~\cite{curtarolo:art107} enables the inexpensive and efficient prediction of the spinodal curve, a quantity which is not usually calculated.
The calculated phase diagrams in Figure~\ref{fig1}(d,e,f) show that TlBiS$_2$-TlBiSe$_2$, TlBiSe$_2$-TlBiTe$_2$, and TlBiS$_2$-TlBiTe$_2$ systems have an asymmetric miscibility gap.
The asymmetric behavior of the miscibility gap is typical when the substitution atoms of the two end members have very different atomic radii~\cite{Burton_jap_2006}.

The calculations indicate that TlBiS$_2$ and TlBiSe$_2$ are miscible above 162K (Figure~\ref{fig1}(d)).
To the best of our knowledge, for the TlBiS$_2$-TlBiSe$_2$ system, there are no experimentally obtained phase diagram data available below 930K,
{
at which temperature 
}
both compounds are still completely miscible~\cite{Babanly_amchemscij_2016}.
Successful high temperature synthesis followed by cooling to room temperature has been reported for several compositions~\cite{Xu_science_2011,Sato_np_2011,Souma_prl_2012}.
For these experiments, homogeneity for $x$=0.8 is indicated at the 8 $\mu$m scale by electron probe microanalysis, although it is unclear at which temperature the analysis was performed~\cite{Sato_np_2011}.
Angle-resolved photoemission spectroscopy ({\small ARPES}) measurements to investigate the novel electronic properties are performed at very low temperatures, where no thermodynamic data is available.
The calculated phase diagram (Figure~\ref{fig1}(d)) indicates that TlBiS$_2$ and TlBiSe$_2$ {should} start to spinodally decompose when the system is cooled down for {\small ARPES} measurements.
{
However, due to slow kinetics at low temperatures, the sample may not have sufficient time to complete the decomposition process prior to the electronic structure measurements.
Without reaching equilibrium, different time-temperature profiles will lead to different experimental results.
Therefore, depending on the details of the experimental procedure, the samples that the {\small ARPES} measurements are performed on could be at different stages of phase separation, which would explain the different results obtained in the two independent electronic structure studies~\cite{Xu_science_2011,Sato_np_2011}.
}

{
For TlBiSe$_2$-TlBiTe$_2$ there are no experimental data available below 760K~\cite{Babanly_amchemscij_2016}.
The Monte Carlo simulations show that TlBiSe$_2$ and TlBiTe$_2$ become immiscible below 400K (Figure \ref{fig1}(e)).
This system is suitable for the practical realization of self-organized heterostructures:
It is possible to obtain a homogeneous stable phase around 500K (well below the melting point of the materials); and to obtain self-organized heterostructures between chemically distinct but iso-structural phases at room temperature.
}

{
  The predicted results show that the TlBiS$_2$-TlBiTe$_2$ system is immiscible below 1040K.
  Reported experimental results indicate a eutectic isotherm at
  $T_{\mathrm{e}}\sim$810K~\cite{Jafarov_im_2014,Babanly_amchemscij_2016}. 
  Its range of composition 
  is depicted in Figure \ref{fig1} as a horizontal line.
  There are two scenarios.
  {\bf i.}
  Any mixture starting outside the eutectic {\it solvi}
  range [TlBi(S$_2$-rich,Te$_2$-poor) or
  TlBi(S$_2$-poor,Te$_2$-rich)] and solidifying through the miscibility gap will not undergo
  eutectic transformation and eventually will find the spinodal line at
  lower temperature.
  {\bf ii.}
  Any composition within the eutectic range will allow a fraction of the
  liquid to reach $x_{\mathrm{e}}\sim$64$\%$ causing a eutectic
  transformation. 
  This is an interesting regime: the potentially sudden atomic
  reorganization, the release of latent heat, and the concomitant
  entrance in the spinodal region,
  will promote the formation of interesting phase-separating
  2D microstructures which can be further tailored through heat
  treatment (eutectic superlattices). 
  Such organizations could have surprising electronic properties. 
  Experiments at and around the eutectic composition with different
  time-temperature profiles are highly suggested.
  Compositions
  slightly above the minimum and 
  slightly below the maximum
  eutectic range could also lead to interesting properties. 
}

Since excess vibrational free energy is neglected in our calculations, 
lower $T_{\mathrm c}$ values can be expected in experiment, 
and similar behavior is reported for other systems~\cite{Adjaoud_prb_2009,Burton_jap_2006}.
{
  Also, as a result of the slow kinetics at low temperatures,
  experimental proof of the miscibility gap has not been reported yet for
  these systems.
}

\begin{figure*}[]%
  \centering
  \includegraphics*[width=0.99\textwidth,clip=true]{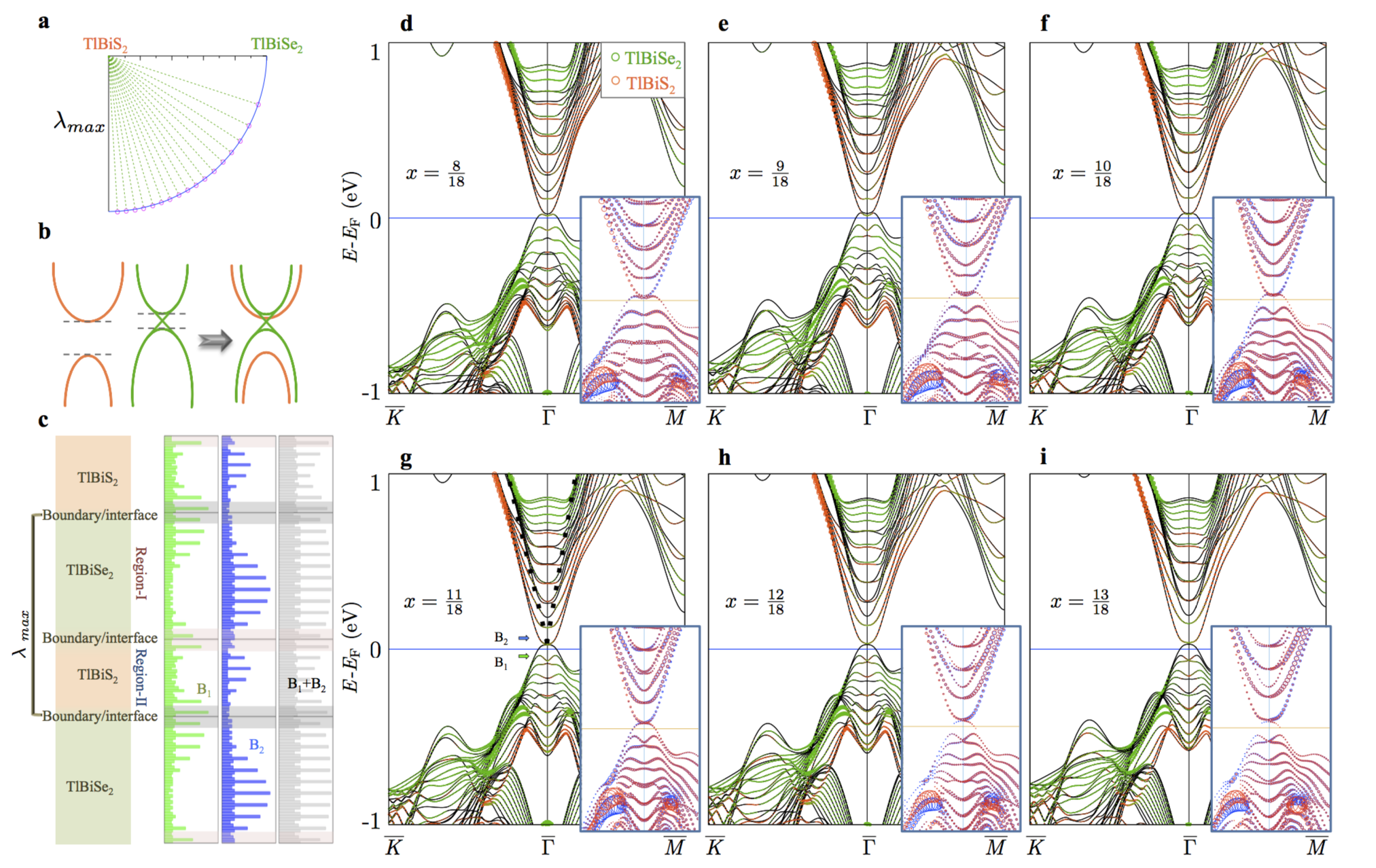}
  \vspace{-2mm}
  \caption{\small
    {\bf (a)} Schematic representation of considered compositions and
    {\bf (b)} band alignment for (TlBiS$_2$)$_{1-x}$(TlBiSe$_2$)$_x$.
    {\bf (c)} Spatial distribution of the charge density {of states B$_1$ and B$_2$ at $\overline \Gamma$ for $x={11}/{18}$.
      The summed charge density of both states is also shown, and exhibits a symmetric shape over the whole system.} 
    Electronic band structure of (TlBiS$_2$)$_{1-x}$(TlBiSe$_2$)$_x$ boundaries with projection of four atomic layers from each side of the boundaries for compositions
    {\bf (d)} $x={8}/{18}$, 
    {\bf (e)} $x={9}/{18}$,
    {\bf (f)} $x={10}/{18}$,
    {\bf (g)} $x={11}/{18}$, 
    {\bf (h)} $x={12}/{18}$ and 
    {\bf (i)} $x={13}/{18}$.
    The insets show the band structure with in-plane spin contributions. 
    The size of symbols are proportional to the degree of in-plane spin polarization in the five atomic layers nearest the interface, and the red and blue colors correspond to the in-plane spin-up and spin-down directions, respectively.
    $\overline K - \overline \Gamma - \overline M$ represents the path in the two-dimensional Brillouin zone of the rhombohedral lattice.}
  \label{fig2}
\end{figure*}

The calculations show that larger mismatches between the size of the interchangeable atoms corresponds to higher $T_{\mathrm c}$ values for these three systems.
Similar behavior has been observed for various systems, such as refractory carbide solid solutions~\cite{Adjaoud_prb_2009}, lead chalcogenides~\cite{curtarolo:art107} and some carbonate quasi-binary systems~\cite{Liu_chemgeo_2016}. \\ \ \\


\noindent{\bf Boundaries of immiscible systems.}
Boundaries of immiscible systems are modelled by keeping the phase boundary wavelength, $\lambda_\mathrm{max}$, constant.
Generally, $\lambda_\mathrm{max}$ ranges from 10 to 100 nm for various systems~\cite{Gelbstein_CHEMMAT_Spinodal_2010,Kanai_JACS_Spinodal_2004,deLaFiguera_PRL_Spinodal_2008}.
For {these} calculations, $\lambda_\mathrm{max}$ is set to 13-14 nm for computational feasibility.
This value also provides sufficient thickness for investigating gapless metallic states~\cite{Chang_prb_2011,Singh_prb_2012}.

In the crystalline solution, finite thickness regions form and interfaces emerge in between them.
Biaxial strain of opposite sign is imposed on both sides of the interface until they reach the mutual lattice parameter. 
Therefore, the strain is shared between each side of the interface (one compressive and the other tensile) so that the misfit between the lattices is reduced. 
For small strains, the system remains coherent. 
However, in the case of large strain, the system interfaces can become semi-coherent with periodically repeating misfit dislocations. 
In epitaxially grown semiconductor heterostructures, the misfit can be accommodated by uniform elastic strain for films below a critical thickness~\cite{Strained_Layer_Superlattices_32,Strained_Layer_Superlattices_33}. 
Analogously, we propose that coherent interfaces can be obtained in the spinodally decomposed systems, even for big lattice misfits, by keeping one region thinner than the critical thickness. 
Therefore, the investigation of the large lattice mismatch systems TlBiSe$_2$/TlBiTe$_2$ and TlBiS$_2$/TlBiTe$_2$ is performed by focusing around the end compositions where one region is very thin. 
For the moderate lattice mismatch system TlBiS$_2$/TlBiSe$_2$, the whole compositional range is investigated. 

\begin{figure*}[]%
  \includegraphics*[width=0.99\textwidth,clip=true]{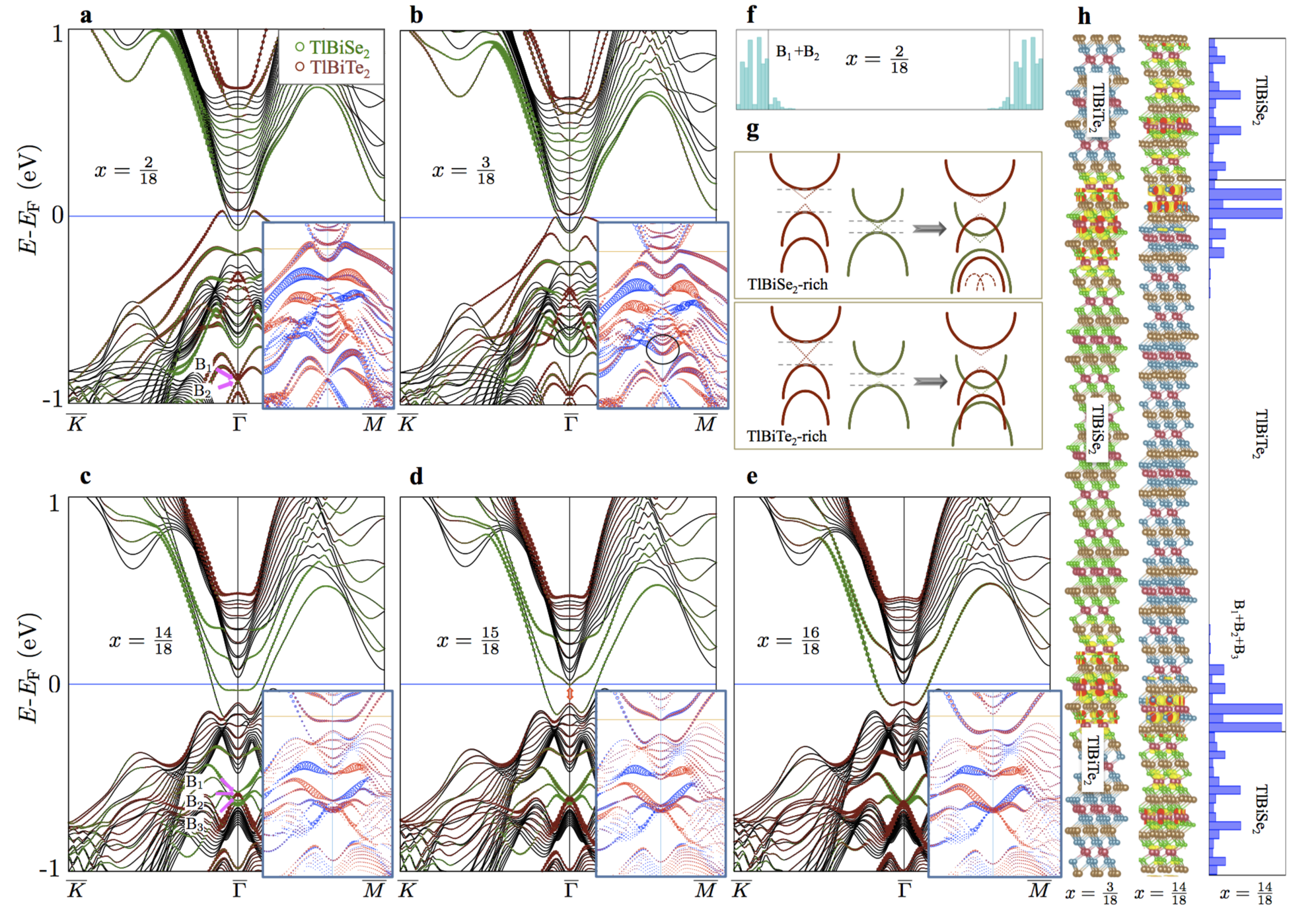}
  \vspace{-2mm}
  \caption{\small Electronic band structure of (TlBiSe$_2$)$_{1-x}$(TlBiTe$_2$)$_x$ boundaries for compositions 
    {\bf (a)} $x={2}/{18}$,
    {\bf (b)} $x={3}/{18}$, 
    {\bf (c)} $x={14}/{18}$, 
    {\bf (d)} $x={15}/{18}$, and 
    {\bf (e)} $x={16}/{18}$.
    The contribution of four atomic layers from each side of the region, or the contribution of the entire thin region (in the case of very thin regions), are projected onto the electronic band structures.
    {\bf (f)} and 
    {\bf (h)} Projected and isosurface representation of {summed} charge density for $x={2}/{18}$, ${3}/{18}$, and ${14}/{18}$.
    {\bf (g)} Schematic representation of band alignment of (TlBiSe$_2$)$_{1-x}$(TlBiTe$_2$)$_x$ boundaries for TlBiSe$_2$-rich and TlBiTe$_2$-rich compositions.
    The insets show the band structure with in-plane spin contributions. 
    The size of symbols are proportional to the degree of in-plane spin polarization in the five atomic layers nearest the interface, and the red and blue colors correspond to the in-plane spin-up and spin-down directions, respectively
    (color code: \textcolor{pranab_red}{$\bullet$}Bi~~~\textcolor{brown}{$\bullet$}Tl~~~\textcolor{green}{$\bullet$}Se~~~\textcolor{PaleBlue}{$\bullet$}Te).
  }
  \label{fig3}
\end{figure*}

If the band inversion takes place at $\Gamma$ in 3D, then the Dirac points will always appear at $\Gamma$ in 2D, independent of the lattice direction~\cite{curtarolo:art77}.
Therefore, the superlattices are modelled along the (111) direction of the rhombohedral cell for computational feasibility, while still capturing the relevant physics.
Creating boundaries between iso-structural and iso-valent phases eliminates the effect of dangling bonds on the electronic structure due to the continuous nature of the system.
For these calculations, the boundaries are placed between the weakly bonded interfacial $X$-Tl layers, although these layers are still considered as strongly interacting due to their ionic/covalent bonding.
This is in contrast to Bi$_2$Se$_3$ and Bi$_2$Te$_3$, which have weak van der Waals bonds between quintuple layers.

The superlattices are modelled by considering the 0K configurations, since the sample cleaving and the {\small ARPES} measurements are mostly carried out at low temperatures ($\sim$ 15-30K)~\cite{Xu_science_2011,Kuroda10prl}.
At 0K, only two separate phases can exist; an interface forms between TlBiS$_2$ and TlBiSe$_2$, TlBiSe$_2$ and TlBiTe$_2$, or TlBiS$_2$ and TlBiTe$_2$.

The band alignment needs to be investigated to define the position of the gapless metallic states and to understand the interaction between the constituent materials bands.
First, the average electrostatic potential difference between the two constituent materials is obtained. 
Next, the difference between the valence band edge and the average electrostatic potential of each constituent material (which is under the same biaxial strain as in the superlattice) is calculated.
For each system, the band alignment is investigated for two compositions.
The results show that although the band offset magnitude changes, the band alignment character of the system is preserved.
This demonstrates the feasibility of engineering the bands of spinodally decomposed materials through compositional tuning. \\


\noindent{\bf (TlBiS$_2$)$_{1-x}$(TlBiSe$_2$)$_x$ boundaries.}
These boundaries represent a possible trivial-insulator/TI system.
In order to find the composition(s) where 2D-gapless metallic states arise, a wide range of composition space is scanned (Figure~\ref{fig2}(a)) while adhering to a constant value of $\lambda_\mathrm{max}$.

The position of the interface states are illustrated by projecting the contribution of four atomic layers from each side of the boundaries.
As can be seen in Figure~\ref{fig2}, metallic states emerge for the ${8}/{18}$ $\le$ $x$ $\le$ ${13}/{18}$ compositional range and the system has a tiny band gap (Figure~\ref{fig2}(d-i)), which is smaller than the room temperature thermal energy (0.025 eV), {and is within the DFT error range}.
The atom resolved band structures indicate that the lower branch of the metallic state originates from TlBiSe$_2$ and the upper branch is from TlBiS$_2$.
The system has staggered (type-II) band alignment (Figure~\ref{fig2}(b)), where the TlBiS$_2$ conduction band and the upper branch of the TlBiSe$_2$ Dirac cone fall in the same energy range.
In fact, the upper branch of the Dirac cone is embedded in the TlBiS$_2$ conduction bands and the biggest contribution comes from TlBiSe$_2$ at $\overline \Gamma$.

\begin{figure*}[]%
\centering
\includegraphics*[width=0.99\textwidth,clip=true]{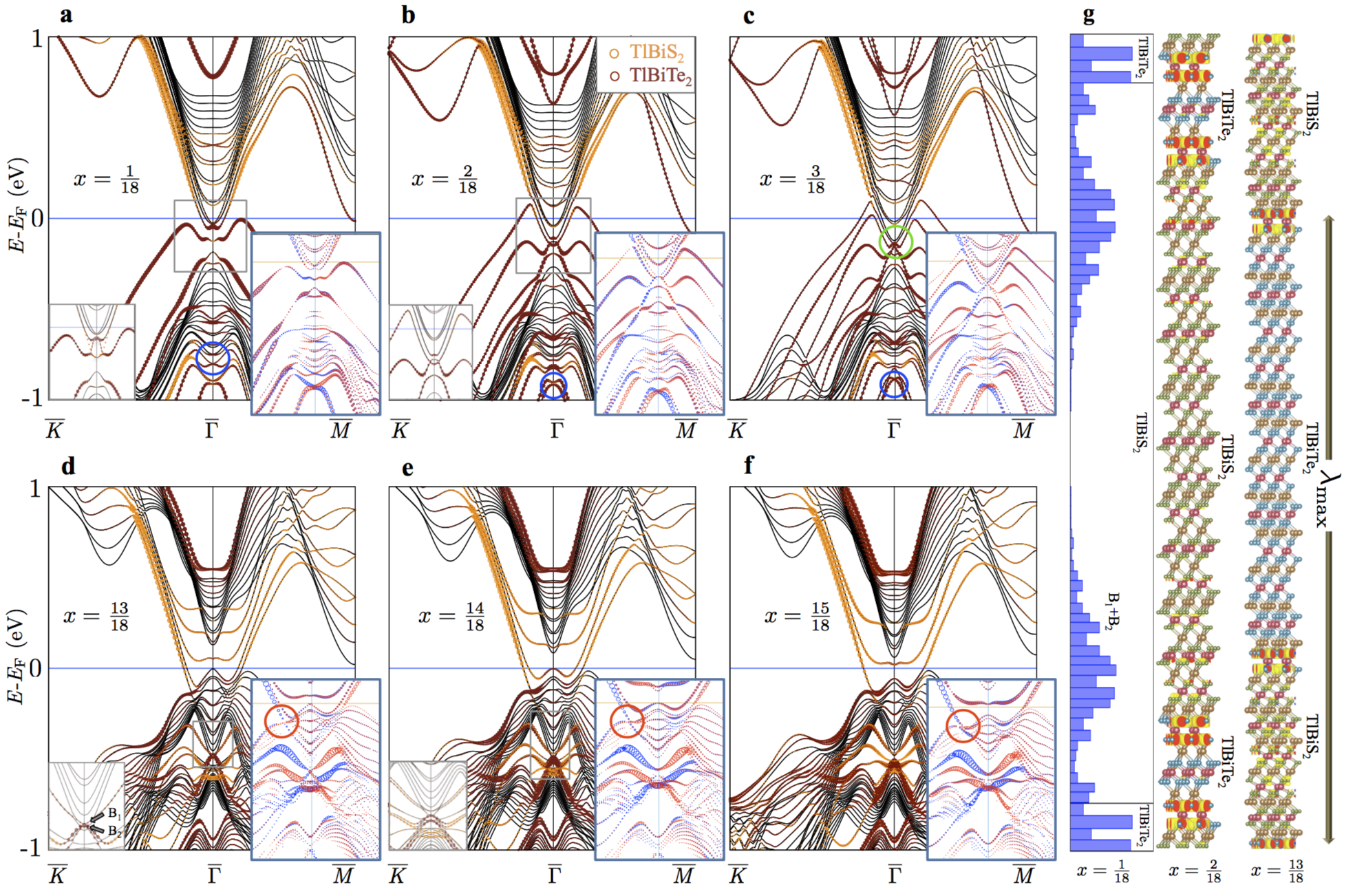}
\vspace{-2mm}
\caption{\small
 Electronic band structure of (TlBiS$_2$)$_{1-x}$(TlBiTe$_2$)$_x$ boundaries for compositions 
{\bf (a)} $x={1}/{18}$, 
{\bf (b)} $x={2}/{18}$, 
{\bf (c)} $x={3}/{18}$, 
{\bf (d)} $x={13}/{18}$, 
{\bf (e)} $x={14}/{18}$, and 
{\bf (f)} $x={15}/{18}$.
The contribution of four atomic layers from each side of the region, or the contribution of the entire thin region (in the case of very thin regions), are projected onto the electronic band structures.
{\bf (g)} Projected and isosurface representation of {summed} charge density {at the Dirac point} for $x={1}/{18}$, $x={2}/{18}$ and $x={14}/{18}$.
The insets show the band structure with in-plane spin contributions. 
The size of symbols are proportional to the degree of in-plane spin polarization in the five atomic layers nearest the interface, and the red and blue colors correspond to the in-plane spin-up and spin-down directions, respectively
(color code: \textcolor{pranab_red}{$\bullet$}Bi~~~\textcolor{brown}{$\bullet$}Tl~~~\textcolor{pranab_green}{$\bullet$}S~~~\textcolor{PaleBlue}{$\bullet$}Te).
}
\label{fig4}
\end{figure*}

Analyzing the projected charge density of the metallic states can provide a better spatial understanding of the electronic structure of the system.
This projection is presented in Figure~\ref{fig2}(c) for composition $x={11}/{18}$ at $\overline \Gamma$.
{B$_1$ and B$_2$ represent the highest and lowest points of the bands which form the Dirac point at $\overline \Gamma$.}
For B$_1$, the charge density accumulates around the interfaces with some penetration through the interior layers.
This agrees well with the reported studies which indicate that the charge density of the surface states decays much slower for TlBi$X_2$, specifically for TlBiSe$_2$, than for Bi$_2$Se$_3$ and Bi$_2$Te$_3$~\cite{Chang_prb_2011,Eremeev_jetpl_2010,Eremeev_prb_2011}.
For the B$_2$ state, the charge density shifts away from the interface as a result of the interaction between the TlBiSe$_2$ Dirac cone and the TlBiS$_2$ conduction band minimum.
This is consistent with previous studies of \TI\ heterostructures, which show that depending on the interaction strength at the interface, the spatial position of the Dirac states can shift away from the interface~\cite{Zhang_acsnano_2012,Jin_prb_2016,Wu_scirep_2013}.
The orbital character of the B$_1$ and B$_2$ bands at $\overline \Gamma$ are mainly $p_z$, consistent with surface states of TlBiSe$_2$, and different from the $p_{xy}$ orbital character of the valence band maximum and conduction band minimum of the bulk (see \SI).
Due to the spatial shift of the B$_2$ state, the total charge density at the Dirac-like point is distributed all over the material instead of accumulating at the interfaces ({Figure~\ref{fig2}(c)}).
{It can, however, be expected that for larger superlattices with thicker TlBi$X_2$ regions (larger $\lambda_\mathrm{max}$), the states will be localized close to the interface.}
The spin polarized band structure (Figure~\ref{fig2}(d-i), insets) illustrates that spin degeneracy is lifted for the upper branch of the gapless metallic states, which is from the trivial insulator (TlBiS$_2$), specifically for $x$=${10}/{18}$ and ${11}/{18}$.
This behavior is similar to the recent studies which report that in semiconductor/TI heterostructures, semiconductor states can acquire a nontrivial spin texture as a result of their interaction with the \TI\ interface states~\cite{Hutasoit_prb_2011,Seixas_nc_2015}. \\


\noindent{\bf (TlBiSe$_2$)$_{1-x}$(TlBiTe$_2$)$_x$ boundaries.}
This system contains two 3D \TI\ materials.
The presence of gapless topological interface states at the junction of two \TIs\ depends on the helicity and magnitude of the Fermi velocity of the \TI\ surface states, and on the mirror symmetry of the system~\cite{Takahashi_prl_2011,Beule_prb_2013}.
Due to the focus around the end compositions at the TlBiSe$_2$(TlBiTe$_2$)-rich part, the TlBiTe$_2$(TlBiSe$_2$) region becomes quite thin, behaving like an ultra thin film.
In addition, the systems have broken band alignment, where the fundamental band gaps of the two constituent materials do not overlap.
No interaction is expected between any present TlBiSe$_2$ and TlBiTe$_2$ metallic {interface} states.
However, due to the narrow band gap of the systems, the overlap between the electron and hole states of TlBiSe$_2$ and TlBiTe$_2$ is large.
The contribution of four atomic layers from each side of the region or the contribution of the entire thin region (in the case of very thin regions) are projected onto the electronic band structures.
This illustrates how changing the composition affects the interface states.

The TlBi$X_2$ systems have ionic/covalent bonding rather than van der Waals bonding between the layers.
Interlayer separation changes with respect to bulk at the interface between TlBiSe$_2$ and TlBiTe$_2$ are less than 0.2 \AA, indicating that the interaction at the interface remains strong.
As a result of this strong interaction, reshaping of the bands is expected.
This is different than the simple superposition of bands where interfacial coupling is weak.
The band structures for various compositions are presented in Figure~\ref{fig3}.
It can be clearly seen that around the two end compositions, one region is thick enough to reproduce the bulk band features.
Since the system is periodically repeating and the vacuum interface is neglected, the Dirac cone from the surface states does not show up in the electronic band structure.
Therefore, in contrast to the \TI\ heterostructures where vacuum interfaces are considered, multiple Dirac cones are not expected here~\cite{Seixas_nc_2015,Jin_prb_2016}.

For TlBiSe$_2$-rich compositions ($x$=${2}/{18}$ and ${3}/{18}$), the TlBiSe$_2$ region is much wider than TlBiTe$_2$, and wider than the theoretically defined critical thickness for the emergence of the Dirac cone.
For these compositions the upper branch of the Dirac cone is present near the Fermi level.
The lower branch is suppressed by TlBiTe$_2$ bulk states, and no gapless states are present.
At $x={2}/{18}$, X-shaped gapless metallic states emerge around -0.85 eV, and these states are mainly from the TlBiTe$_2$ region and mostly localized at the {interface} (Figure ~\ref{fig3}(f)).
However, they are topologically trivial states since they have spin degeneracy.
A similar X-shaped band is reported for one bilayer Bi on the TlBiSe$_2$ surface~\cite{Shoman_nc_2015}. 
At $x={3}/{18}$ (Figure~\ref{fig3}(b)), gapless {interface} states arise around -0.7 eV at $\overline \Gamma$ (black ellipse).
The isosurface charge density indicates that these states are TlBiSe$_2$ {interface} states and have $p_z$ character.
Also, the spin resolved band structure shows that these bands are spin polarized (Figure~\ref{fig3}(b), inset). 
A similar state is reported for other \TI\ superlattices~\cite{Johannsen_prb_2015,Gibson_prb_2013,Weber_prl_2015,Eschbach_nc_2017}.

At the TlBiTe$_2$-rich compositions, the thickness and relative strains on the regions change.
This does not alter the band alignment character, only tunes the band offset.
For these compositions, the TlBiTe$_2$ region is thick enough to obtain a Dirac cone from {interface} states, but the {interface} states are gapped near the Fermi level, as indicated by the red arrow (Figure~\ref{fig3}(d)).
However, Dirac-like gapless states emerge between the TlBiSe$_2$ and TlBiTe$_2$ states around -0.5 eV.
In order to understand the nature of these states, the projected charge density as well as the isosurface charge density at $\overline \Gamma$ is presented in Figure ~\ref{fig3}(h) for $x={14}/{18}$.
The plots indicate that these Dirac-like points are spread across the TlBiTe$_2$ {interface} state and the TlBiSe$_2$ bulk state.
These types of bands are reported for \TI\ heterostructures and are described as mixed-character bands~\cite{Seixas_nc_2015,Jin_prb_2016}.
Thusly, instead of two metallic channels, only one metallic channel arises, which includes the entire TlBiSe$_2$ region.
The upper and lower branches of the Dirac-like points have opposite spin polarization (Figure~\ref{fig3}(c-e), insets), indicating that they are topologically protected. \\

\noindent{\bf (TlBiS$_2$)$_{1-x}$(TlBiTe$_2$)$_x$ boundaries.}
This system is another example of a possible trivial-insulator/TI system with a large lattice misfit.

In the TlBiS$_2$-rich compositions, the TlBiTe$_2$ region is very thin, so the system is a superlattice of a trivial insulator and an ultra thin film of 3D \TI\ with broken band alignment.
The band structures in Figure~\ref{fig4}(a-c) demonstrate that at compositions $x ={1}/{18}$, ${2}/{18}$, and ${3}/{18}$, there is band inversion between the TlBiS$_2$ and TlBiTe$_2$ states, and their hybridization causes the gap to open near $\overline \Gamma$.
This indicates the non-trivial nature of the system.
In addition, for $x ={1}/{18}$ a Dirac-like point appears around -0.8 eV (blue circle) and the projected charge density (Figure~\ref{fig4}(g), left) demonstrates that charges accumulate close to the TlBiS$_2$ edge, with some shift due to the interaction with the TlBiTe$_2$ states, along with some contribution from the TlBiTe$_2$ region.
The spin texture indicates that these states are spin polarized (Figure~\ref{fig4}(a), inset).
Increasing the composition ($x$) tunes the relative thickness of the TlBiS$_2$ and TlBiTe$_2$ regions, as well as the strains on them, changing the population and position of the states.
Inverted bands exist around $\overline \Gamma$ for the TlBiS$_2$-rich compositions.
In addition, Rashba-like states emerge around -0.9 eV (Figure~\ref{fig4}(b,c)).
As can be seen from Figure~\ref{fig4}(c), at $x ={3}/{18}$ a Dirac point emerges in the inverted gap near the Fermi level.
The spatial distribution of the Dirac point charge density shows that charge accumulates around the interface, mainly at the edge of the TlBiTe$_2$ region.
In contrast to the surface Dirac cones, these states have $p_{xy}$ orbital character.
The spin texture of the Dirac cone is presented in Figure~\ref{fig4}(c), inset.

At the TlBiTe$_2$-rich region, (TlBiS$_2$)$_{1-x}$(TlBiTe$_2$)$_x$ and (TlBiSe$_2$)$_{1-x}$(TlBiTe$_2$)$_x$ demonstrate very similar band dispersion, since both systems have broken band alignment and common constituent materials.
Replacing TlBiSe$_2$ with TlBiS$_2$ does not make a significant change in the band dispersion, although the Se atom has a bigger spin-orbit coupling constant than the S atom.
However, the spin texture is changed for the Dirac-like states.
At $x ={13}/{18}$ (Figure~\ref{fig4}(d)), a 2D character Dirac-like point arises around -0.5 eV and the corresponding charge density is localized at the interface, mostly on the TlBiTe$_2$ side (Figure~\ref{fig4}(g), right).
However, the spin texture (Figure~\ref{fig4}(d-f), inset) indicates that these bands do not have opposite spin direction for the upper and lower branches.
Therefore, they are trivial metallic interface states.
In addition, the TlBiS$_2$ and TlBiTe$_2$ bands cross each other away from the time reversal invariant momentum ({\small TRIM}) points along the $\overline K$-$\overline \Gamma$ path direction, and this crossing happens between opposite spin bands (Figure~\ref{fig4}(d-f), red circle).
Recently, similar bands have been reported for Bi$_4$Se$_3$ and Bi$_1$Te$_1$ superlattices~\cite{Weber_prl_2015, Eschbach_nc_2017}.
In contrast to our findings, their results showed that the band crossing along the $\overline M$-$\overline \Gamma$ direction is gapless and mirror symmetry protected, and also that another band crossing happens at $\overline \Gamma$: the state is dually protected.


\section*{\label{sec:conc}Conclusion}
In this work, the electronic properties of thermodynamically formed topological insulator superlattices have been investigated.
Our results demonstrate that it is possible to obtain self-assembled interfaces between iso-structural and iso-valent materials with interesting electronic properties.
Compositional tuning induces various phenomena, such as topological interface states, spin texture gain by non-topological states, band inversion, band crossing between the {\small TRIM} points, and Rashba-like states.
The emergence of these phenomena is related to the band alignment and composition of the systems.
With this method, we demonstrate that obtaining 2D metallic channels in a 3D insulating matrix is possible for \TI\ superlattices without vacuum interfaces.
Since these states are protected from the environment, they can be useful for device applications.
These findings suggest that the combination of thermodynamic and electronic properties can create a pathway for investigating thermodynamically driven interfaces to obtain possible novel phenomena, which may help to explore more materials for spintronic devices and further applications.
 
\section*{Acknowledgments}
The authors acknowledge support by DOD-ONR (N00014-13-1-0635, N00014-15-1-2863, N00014-16-1-2326).
The consortium \AFLOW.org acknowledges Duke University -- Center for Materials Genomics -- for computational support.
S.C. acknowledges the Alexander von Humboldt Foundation for financial support.
The authors thank Corey Oses, David Hicks, Eric Gossett, and Ohad Levy for helpful discussions.

\section*{Supporting Information}
Optimized lattice parameters and atom resolved band structures of rhombohedral TlBi$X_2$ ($X$=S, Se, Te) bulk systems;
atom resolved surface electronic structure of $X$-terminated and Tl-terminated (111) surfaces of TlBi$X_2$ ($X$=S, Se, Te).

\begin{thebibliography}{10}
\expandafter\ifx\csname urlstyle\endcsname\relax
  \providecommand{\doi}[1]{doi:\discretionary{}{}{}#1}\else
  \providecommand{\doi}{doi:\discretionary{}{}{}\begingroup
  \urlstyle{rm}\Url}\fi
\providecommand{\selectlanguage}[1]{\relax}
\providecommand{\bibAnnoteFile}[1]{%
  \IfFileExists{#1}{\begin{quotation}\noindent\textsc{Key:} #1\\
  \textsc{Annotation:}\ \input{#1}\end{quotation}}{}}
\providecommand{\bibAnnote}[2]{%
  \begin{quotation}\noindent\textsc{Key:} #1\\
  \textsc{Annotation:}\ #2\end{quotation}}

\bibitem{Cressler_2003}
J.~D. Cressler and G.~Niu, \emph{Silicon-germanium Heterojunction {Bi}polar
  Transistors} (Artech House, London, Boston, 2003).
\input{Cressler_2003}

\bibitem{Razeghi_jap_1996}
M.~Razeghi and A.~Rogalski, \emph{Semiconductor ultraviolet detectors}, J.\
  Appl.\ Phys. \textbf{79}, 7433 (1996).
\input{Razeghi_jap_1996}

\bibitem{Rizzi_jas_2010}
M.~Rizzi, M.~D'{A}loia, and B.~Castagnolo, \emph{Semiconductor Detectors and
  Principles of Radiation-matter Interaction}, J.\ Appl.\ Sci. \textbf{10},
  3141--3155 (2010).
\input{Rizzi_jas_2010}

\bibitem{Hwang_apl_2005}
D.-K. Hwang, S.-H. Kang, J.-H. Lim, E.-J. Yang, J.-Y. Oh, J.-H. Yang, and S.-J.
  Park, \emph{$p$-{Zn}{O}/$n$-{Ga}{N} heterostructure {Zn}{O} light-emitting
  diodes}, Appl.\ Phys.\ Lett. \textbf{86}, 222101 (2005).
\input{Hwang_apl_2005}

\bibitem{Kohler_nature_2002}
R.~K\"{o}hler, A.~Tredicucci, F.~Beltram, H.~E. Beere, E.~H. Linfield, A.~G.
  Davies, D.~A. Ritchie, R.~C. Iotti, and F.~Rossi, \emph{Terahertz
  semiconductor-heterostructure laser}, Nature \textbf{417}, 156–159 (2002).
\input{Kohler_nature_2002}

\bibitem{curtarolo:art81}
S.~Curtarolo, G.~L.~W. Hart, M.~{Buongiorno Nardelli}, N.~Mingo, S.~Sanvito,
  and O.~Levy, \emph{The high-throughput highway to computational materials
  design}, Nat.\ Mater. \textbf{12}, 191--201 (2013).
\input{curtarolo:art81}

\bibitem{exarhos_discovery-based_2007}
G.~J. Exarhos and X.-D. Zhou, \emph{Discovery-based design of transparent
  conducting oxide films}, Thin\ Solid\ Films \textbf{515}, 7025--7052 (2007).
\input{exarhos_discovery-based_2007}

\bibitem{curtarolo:art97}
E.~Sachet, C.~T. Shelton, J.~S. Harris, B.~E. Gaddy, D.~L. Irving,
  S.~Curtarolo, B.~F. Donovan, P.~E. Hopkins, P.~A. Sharma, A.~L. Sharma,
  J.~Ihlefeld, S.~Franzen, and J.-P. Maria, \emph{Dysprosium-doped cadmium
  oxide as a gateway material for mid-infrared plasmonics}, Nat.\ Mater.
  \textbf{14}, 414--420 (2015).
\input{curtarolo:art97}

\bibitem{Nakamura_NMAT_materials_scarcity_2011}
E.~Nakamura and K.~Sato, \emph{Managing the scarcity of chemical elements},
  Nat.\ Mater. \textbf{10}, 158--161 (2011).
\input{Nakamura_NMAT_materials_scarcity_2011}

\bibitem{curtarolo:art107}
D.~Usanmaz, P.~Nath, J.~J. Plata, G.~L.~W. Hart, I.~Takeuchi, M.~{Buongiorno
  Nardelli}, M.~Fornari, and S.~Curtarolo, \emph{First principles
  thermodynamical modeling of the binodal and spinodal curves in lead
  chalcogenides}, Phys.\ Chem.\ Chem.\ Phys. \textbf{18}, 5005--5011 (2016).
\input{curtarolo:art107}

\bibitem{Hirahara_prl_2011}
T.~Hirahara, G.~{Bi}hlmayer, Y.~Sakamoto, M.~Yamada, H.~Miyazaki, S.-I. Kimura,
  S.~Bl\"{u}gel, and S.~Hasegawa, \emph{Interfacing 2D and 3D Topological
  Insulators: {Bi}(111) {Bi}layer on {Bi}$_2${Te}$_3$}, Phys.\ Rev.\ Lett.
  \textbf{107}, 166801 (2011).
\input{Hirahara_prl_2011}

\bibitem{Berntsen_prb_1013}
M.~H. Berntsen, O.~G\"{o}tberg, B.~M. Wojek, and O.~Tjernberg, \emph{Direct
  observation of decoupled Dirac states at the interface between topological
  and normal insulators}, Phys.\ Rev.\ B \textbf{88}, 195132 (2013).
\input{Berntsen_prb_1013}

\bibitem{Shoman_nc_2015}
T.~Shoman, A.~Takayama, T.~Sato, S.~Souma, T.~Takahashi, T.~Oguchi,
  K.~{Se}gawa, and Y.~Ando, \emph{Topological proximity effect in a topological
  insulator hybrid}, Nat.\ Commun. \textbf{6}, 6547 (2015).
\input{Shoman_nc_2015}

\bibitem{Yan_rpp_2012}
B.~Yan and S.-C. Zhang, \emph{Topological materials}, Rep.\ Prog.\ Phys.
  \textbf{75}, 096501 (2012).
\input{Yan_rpp_2012}

\bibitem{Chen_science_2009}
Y.~L. Chen, J.~G. Analytis, J.-H. Chu, Z.~K. Liu, S.-K. Mo, X.~L. Qi, H.~J.
  Zhang, D.~H. Lu, X.~Dai, Z.~Fang, S.~C. Zhang, I.~R. Fisher, Z.~Hussain, and
  Z.-X. Shen, \emph{Experimental Realization of a Three-Dimensional Topological
  Insulator, {Bi}$_2${Te}$_3$}, Science \textbf{325}, 178--181 (2009).
\input{Chen_science_2009}

\bibitem{Zhang_Nat.Phys._2009}
H.~Zhang, C.-X. Liu, X.-L. Qi, X.~Dai, Z.~Fang, and S.-C. Zhang,
  \emph{Topological insulators in {Bi$_2$Se$_3$}, {Bi$_2$Te$_3$} and
  {Sb$_2$Te$_3$} with a single Dirac cone on the surface}, Nat.\ Phys.
  \textbf{5}, 438--442 (2009).
\input{Zhang_Nat.Phys._2009}

\bibitem{zhang_PRL_2009}
T.~Zhang, P.~Cheng, X.~Chen, J.-F. Jia, X.~Ma, K.~He, L.~Wang, H.~Zhang,
  X.~Dai, Z.~Fang, X.~Xie, and Q.-K. Xue, \emph{Experimental Demonstration of
  Topological Surface States Protected by Time-Reversal Symmetry}, Phys.\ Rev.\
  Lett. \textbf{103}, 266803 (2009).
\input{zhang_PRL_2009}

\bibitem{curtarolo:art77}
K.~Yang, W.~Setyawan, S.~Wang, M.~{Buongiorno Nardelli}, and S.~Curtarolo,
  \emph{A search model for topological insulators with high-throughput
  robustness descriptors}, Nat.\ Mater. \textbf{11}, 614--619 (2012).
\input{curtarolo:art77}

\bibitem{Hsieh_natcom_2012}
T.~H. Hsieh, H.~Lin, J.~Liu, W.~Duan, A.~Bansil, and L.~Fu, \emph{Toplogical
  crystalline insulators in the {Sn}{Te} material class}, Nat.\ Commun.
  \textbf{3}, 982 (2012).
\input{Hsieh_natcom_2012}

\bibitem{Dziawa_NatMAt_2012}
P.~Dziawa, B.~J. Kowalski, K.~Dybko, R.~Buczko, A.~Szczerbakow, M.~Szot,
  E.~Łusakowska, T.~Balasubramanian, B.~M. Wojek, M.~H. Berntsen,
  O.~Tjernberg, and T.~Story, \emph{Topological crystalline insulator states in
  Pb$_{1-x}$Sn$_x$Se}, Nat.\ Mater. \textbf{11}, 1023–1027 (2012).
\input{Dziawa_NatMAt_2012}

\bibitem{Tanaka_NatPhy_2012}
Y.~Tanaka, Z.~Ren, T.~Sato, K.~Nakayama, S.~Souma, T.~Takahashi, K.~Segawa, and
  Y.~Ando, \emph{Experimental realization of a topological crystalline
  insulator in {Sn}{Te}}, Nat.\ Phys. \textbf{8}, 800–803 (2012).
\input{Tanaka_NatPhy_2012}

\bibitem{Ando_jpsj_2013}
Y.~Ando, \emph{Topological insulator materials}, J.\ Phys.\ Soc.\ Jpn.
  \textbf{82}, 102001 (2013).
\input{Ando_jpsj_2013}

\bibitem{Nakayama_PRL_2012}
K.~Nakayama, K.~Eto, Y.~Tanaka, T.~Sato, S.~Souma, T.~Takahashi, K.~Segawa, and
  Y.~Ando, \emph{Manipulation of Topological States and the Bulk Band Gap Using
  Natural Heterostructures of a Topological Insulator}, Phys.\ Rev.\ Lett.
  \textbf{109}, 236804 (2012).
\input{Nakayama_PRL_2012}

\bibitem{Rauch_prb_2013}
T.~Rauch, M.~Flieger, J.~Henk, and I.~Mertig, \emph{Nontrivial interface states
  confined between two topological insulators}, Phys.\ Rev.\ B \textbf{88},
  245120 (2013).
\input{Rauch_prb_2013}

\bibitem{Wu_scirep_2013}
G.~Wu, H.~Chen, Y.~Sun, X.~Li, P.~Cui, C.~Franchini, J.~Wang, X.-Q. Chen, and
  Z.~Zhang, \emph{Tuning the vertical location of helical surface states in
  topological insulator heterostructures via dual-proximity effects}, Sci.\
  Rep. \textbf{3}, 1233 (2013).
\input{Wu_scirep_2013}

\bibitem{Chen_nanolett_2015}
Z.~Chen, L.~Zhao, K.~Park, T.~A. Garcia, M.~C. Tamargo, and L.~Krusin-Elbaum,
  \emph{Robust Topological Interface and Charge Transfer in Epitaxial
  {Bi}$_2${Se}$_3$/{II}-{VI} Semiconductor Superlattices}, Nano\ Lett.
  \textbf{15}, 6365 (2015).
\input{Chen_nanolett_2015}

\bibitem{Xu_science_2011}
S.-Y. Xu, Y.~Xia, L.~A. Wray, S.~Jia, F.~Meier, J.~H. Dil, J.~Osterwalder,
  B.~Slomski, A.~Bansil, H.~Lin, R.~J. Cava, and M.~Z. Hasan, \emph{Topological
  phase transition and texture inversion in a tunable topological insulator},
  Science \textbf{332}, 560 (2011).
\input{Xu_science_2011}

\bibitem{Zhang10natphys}
Y.~Zhang, K.~He, C.-Z. Chang, C.-L. Song, L.-L. Wang, X.~Chen, J.-F. Jia,
  Z.~Fang, X.~Dai, W.-Y. Shan, S.-Q. Shen, Q.~Niu, X.-L. Qi, S.-C. Zhang, X.-C.
  Ma, and Q.-K. Xue, \emph{Crossover of the three-dimensional topological
  insulator {Bi$_2$Se$_3$} to the two-dimensional limit}, Nat.\ Phys.
  \textbf{6}, 584--588 (2010).
\input{Zhang10natphys}

\bibitem{Eschbach_nc_2017}
M.~{\it et al.}. Eschbach, \emph{{Bi}$_1${Te}$_1$ is a dual topological
  insulator}, Nat.\ Commun. \textbf{8}, 14976 (2017).
\input{Eschbach_nc_2017}

\bibitem{Weber_prl_2015}
A.~P. Weber, Q.~D. Gibson, H.~Ji, A.~N. Caruso, A.~V. Fedorov, R.~J. Cava, and
  T.~Valla, \emph{Gapped Surface States in a Strong-Topological-Insulator
  Material}, Phys.\ Rev.\ Lett. \textbf{114}, 256401 (2015).
\input{Weber_prl_2015}

\bibitem{Belopolski_sciadv_2017}
I.~Belopolski, S.-Y. Xu, N.~Koirala, C.~Liu, G.~{Bi}an, V.~N. Strocov,
  G.~Chang, M.~Neupane, N.~Alidoust, D.~Sanchez, H.~Zheng, M.~Brahlek,
  V.~Rogalev, T.~Kim, N.~C. Plumb, C.~Chen, F.~Bertran, P.~{Le F{\'{e}}vre},
  A.~Taleb-Ibrahimi, M.-C. Asensio, M.~Shi, H.~Lin, M.~Hoesch, S.~Oh, and M.~Z.
  Hasan, \emph{A novel artificial condensed matter lattice and a new platform
  for one-dimensional topological phases}, Sci.\ Adv. \textbf{3}, e1501692
  (2017).
\input{Belopolski_sciadv_2017}

\bibitem{Hagmann_njp_2017}
J.~A. Hagmann, X.~Li, S.~Chowdhury, S.-N. Dong, S.~Rouvimov, S.~J.
  Pookpanratana, K.~M. Yu, T.~A. Orlova, T.~B. Bolin, C.~U. Segre, D.~G.
  Seiler, C.~A. Richter, X.~Liu, M.~Dobrowolska, and J.~K. Furdyna,
  \emph{Molecular beam epitaxy growth and structure of self-assembled
  {Bi}$_2${Se}$_3$/{Bi}$_2${Mn}{Se}$_4$ multilayer heterostructures}, New\ J.\
  Phys. \textbf{19}, 085002 (2017).
\input{Hagmann_njp_2017}

\bibitem{Burkov_prl_2011}
A.~A. Burkov and L.~Balents, \emph{{W}eyl Semimetal in a Topological Insulator
  Multilayer}, Phys.\ Rev.\ Lett. \textbf{107}, 127205 (2011).
\input{Burkov_prl_2011}

\bibitem{Androulakis_jacs_2007}
J.~Androulakis, C.-H. Lin, H.-J. Kong, C.~Uher, C.-I. Wu, T.~Hogan, B.~A. Cook,
  T.~Caillat, K.~M. Paraskevopoulos, and M.~G. Kanatzidis, \emph{Spinodal
  Decomposition and Nucleation and Growth as a Means to Bulk Nanostructured
  Thermoelectrics: Enhanced Performance in {Pb}$_{1-x}${Sn}$_x${Te}-{Pb}{S}},
  J.\ Amer.\ Chem.\ Soc. \textbf{129}, 9780--9788 (2007).
\input{Androulakis_jacs_2007}

\bibitem{Zhao_jacs_2012}
L.-D. Zhao, J.~He, S.~Hao, C.-I. Wu, T.~P. Hogan, C.~Wolverton, V.~P. Dravid,
  and M.~G. Kanatzidis, \emph{Raising the thermoelectric performance of p-type
  {Pb}{S} with endotaxial nanostructuring and valence-band offset engineering
  using {Cd}{S} and {Zn}{S}}, J.\ Amer.\ Chem.\ Soc. \textbf{134}, 16327--16336
  (2012).
\input{Zhao_jacs_2012}

\bibitem{Gelbstein_CHEMMAT_Spinodal_2010}
Y.~Gelbstein, B.~Dado, O.~Ben-Yehuda, Y.~Sadia, Z.~Dashevsky, and M.~P. Dariel,
  \emph{High Thermoelectric Figure of Merit and Nanostructuring in Bulk {\it
  p}-type Ge$_x$(Sn$_y$Pb$_{1-y}$)$_{1-x}${Te} Alloys Following a Spinodal
  Decomposition Reaction}, Chem.\ Mater. \textbf{22}, 1054--1058 (2010).
\input{Gelbstein_CHEMMAT_Spinodal_2010}

\bibitem{Sootsman_JAP_2009}
J.~R. Sootsman, J.~He, V.~P. Dravid, C.-P. {Li}, C.~Uher, and M.~G. Kanatzidis,
  \emph{High thermoelectric figure of merit and improved mechanical properties
  in melt quenched PbTe-Ge and PbTe-Ge$_{1-x}$Si$_x$ eutectic and hypereutectic
  composites}, J.\ Appl.\ Phys. \textbf{105}, 083718 (2009).
\input{Sootsman_JAP_2009}

\bibitem{Hillert1961525}
M.~Hillert, \emph{A solid-solution model for inhomogeneous systems}, Acta\
  Metallurgica \textbf{9}, 525--535 (1961).
\input{Hillert1961525}

\bibitem{Cahn1961795}
J.~W. Cahn, \emph{On spinodal decomposition}, Acta\ Metallurgica \textbf{9},
  795--801 (1961).
\input{Cahn1961795}

\bibitem{Cahn1962179}
J.~W. Cahn, \emph{On spinodal decomposition in cubic crystals}, Acta\
  Metallurgica \textbf{10}, 179--183 (1962).
\input{Cahn1962179}

\bibitem{mannhart_oxide_2010}
J.~Mannhart and D.~G. Schlom, \emph{Oxide {Interfaces—An} Opportunity for
  Electronics}, Science \textbf{327}, 1607--1611 (2010).
\input{mannhart_oxide_2010}

\bibitem{yoshimatsu_dimensional-crossover-driven_2010}
K.~Yoshimatsu, T.~Okabe, H.~Kumigashira, S.~Okamoto, S.~Aizaki, A.~Fujimori,
  and M.~Oshima, \emph{Dimensional-Crossover-Driven Metal-Insulator Transition
  in {S}r{V}{O}$_{3}$ Ultrathin Films}, Phys.\ Rev.\ Lett. \textbf{104}, 147601
  (2010).
\input{yoshimatsu_dimensional-crossover-driven_2010}

\bibitem{yoshimatsu_origin_2008}
K.~Yoshimatsu, R.~Yasuhara, H.~Kumigashira, and M.~Oshima, \emph{Origin of
  Metallic States at the Heterointerface between the Band Insulators
  LaAlO$_{3}$ and SrTiO$_{3}$}, Phys.\ Rev.\ Lett. \textbf{101}, 026802 (2008).
\input{yoshimatsu_origin_2008}

\bibitem{Analytis_prb_2010}
J.~G. Analytis, J.-H. Chu, Y.~Chen, F.~Corredor, R.~D. McDonald, Z.~X. Shen,
  and I.~R. Fisher, \emph{Bulk Fermi surface coexistence with Dirac surface
  state in {Bi}$_2${Se}$_3$: A comparison of photoemission and Shubnikov–de
  Haas measurements}, Phys.\ Rev.\ B \textbf{81}, 205407 (2010).
\input{Analytis_prb_2010}

\bibitem{Eto_prb_2010}
K.~Eto, Z.~Ren, A.~A. Taskin, K.~{Se}gawa, and Y.~Ando, \emph{Angular-dependent
  oscillations of the magnetoresistance in {Bi}$_2${Se}$_3$ due to the
  three-dimensional bulk {Fermi} surface}, Phys.\ Rev.\ B \textbf{81}, 195309
  (2010).
\input{Eto_prb_2010}

\bibitem{Butch_prb_2010}
N.~P. Butch, K.~Kirshenbaum, P.~Syers, A.~B. Sushkov, G.~S. Jenkins, H.~D.
  Drew, and J.~Paglione, \emph{Strong surface scattering in ultrahigh-mobility
  {Bi}$_2${Se}$_3$ topological insulator crystals}, Phys.\ Rev.\ B \textbf{81},
  241301(R) (2010).
\input{Butch_prb_2010}

\bibitem{Yan_epl_2010}
B.~Yan, C.-X. Liu, H.-J. Zhang, C.-Y. Yam, X.-L. Qi, T.~Frauenheim, and S.-C.
  Zhang, \emph{Theoretical prediction of topological insulators in
  thallium-based {III}-{V}-{VI}$_2$ ternary chalcogenides}, EPL-Europhys.\
  Lett. \textbf{90}, 37002 (2010).
\input{Yan_epl_2010}

\bibitem{Zhang_sr_2015}
Q.~Zhang, Y.~Cheng, and U.~Schwingenschl\"{o}gl, \emph{Emergence of topological
  and topological crystalline phases in {Tl}{Bi}S$_2$ and {Tl}{Sb}S$_2$}, Sci.\
  Rep. \textbf{5}, 8379 (2015).
\input{Zhang_sr_2015}

\bibitem{Singh_jap_2014}
B.~Singh, H.~Lin, R.~Prasad, and A.~Bansil, \emph{Topological phase transition
  and quantum spin Hall state in {Tl}{Bi}S$_2$}, J.\ Appl.\ Phys. \textbf{116},
  033704 (2014).
\input{Singh_jap_2014}

\bibitem{Sato_np_2011}
T.~Sato, K.~{Se}gawa, K.~Kosaka, S.~Souma, K.~Nakayama, K.~Eto, T.~Minami,
  Y.~Ando, and T.~Takahashi, \emph{Unexpected mass acquisition of Dirac
  fermions at the quantum phase transition of a topological insulator}, Nat.\
  Phys. \textbf{7}, 840--844 (2011).
\input{Sato_np_2011}

\bibitem{Souma_prl_2012}
S.~Souma, M.~Komatsu, M.~Nomura, T.~Sato, A.~Takayama, T.~Takahashi, K.~Eto,
  K.~{Se}gawa, and Y.~Ando, \emph{Spin Polarization of Gapped Dirac Surface
  States Near the Topological Phase Transition in
  {Tl}{Bi}({S}$_{1-x}${Se}$_x$)$_2$}, Phys.\ Rev.\ Lett. \textbf{109}, 186804
  (2012).
\input{Souma_prl_2012}

\bibitem{Singh_prb_2012}
B.~Singh, A.~Sharma, H.~Lin, M.~Z. Hasan, R.~Prasad, and A.~Bansil,
  \emph{Topological electronic structure and {W}eyl semimetal in the
  {Tl}{Bi}{Se}$_2$ class of semiconductors}, Phys.\ Rev.\ B \textbf{86}, 115208
  (2012).
\input{Singh_prb_2012}

\bibitem{Walle_calphad_2002}
A.~van~de Walle, M.~D. Asta, and G.~Ceder, \emph{The alloy theoretic automated
  toolkit: A user guide}, Calphad \textbf{26}, 539--553 (2002).
\input{Walle_calphad_2002}

\bibitem{atat1}
A.~{van~de~Walle} and G.~Ceder, \emph{Automating First-Principles Phase Diagram
  Calculations}, J.\ Phase Equilib. \textbf{23}, 348--359 (2002).
\input{atat1}

\bibitem{axel_MC}
A.~{{v}an {d}e Walle} and M.~D. Asta, \emph{Self-driven lattice-model {Mo}nte
  {Ca}rlo simulations of alloy thermodynamic properties and phase diagrams},
  Model.\ Simul.\ Mater.\ Sci.\ Eng. \textbf{10}, 521 (2002).
\input{axel_MC}

\bibitem{curtarolo:art65}
S.~Curtarolo, W.~Setyawan, G.~L.~W. Hart, M.~Jahn\'{a}tek, R.~V. Chepulskii,
  R.~H. Taylor, S.~Wang, J.~Xue, K.~Yang, O.~Levy, M.~J. Mehl, H.~T. Stokes,
  D.~O. Demchenko, and D.~Morgan, \emph{{AFLOW}: An automatic framework for
  high-throughput materials discovery}, Comput.\ Mater.\ Sci. \textbf{58},
  218--226 (2012).
\input{curtarolo:art65}

\bibitem{curtarolo:art110}
K.~Yang, C.~Oses, and S.~Curtarolo, \emph{Modeling Off-Stoichiometry Materials
  with a High-Throughput {\it Ab-Initio} Approach}, Chem.\ Mater. \textbf{28},
  6484--6492 (2016).
\input{curtarolo:art110}

\bibitem{curtarolo:art104}
C.~E. Calderon, J.~J. Plata, C.~Toher, C.~Oses, O.~Levy, M.~Fornari, A.~Natan,
  M.~J. Mehl, G.~L.~W. Hart, M.~{Buongiorno Nardelli}, and S.~Curtarolo,
  \emph{The {AFLOW} standard for high-throughput materials science
  calculations}, Comput.\ Mater.\ Sci. \textbf{108 Part A}, 233--238 (2015).
\input{curtarolo:art104}

\bibitem{curtarolo:art63}
O.~Levy, M.~Jahn\'{a}tek, R.~V. Chepulskii, G.~L.~W. Hart, and S.~Curtarolo,
  \emph{Ordered Structures in {Rh}enium Binary Alloys from First-Principles
  Calculations}, J.\ Amer.\ Chem.\ Soc. \textbf{133}, 158--163 (2011).
\input{curtarolo:art63}

\bibitem{curtarolo:art57}
O.~Levy, G.~L.~W. Hart, and S.~Curtarolo, \emph{Structure maps for {h}cp metals
  from first-principles calculations}, Phys.\ Rev.\ B \textbf{81}, 174106
  (2010).
\input{curtarolo:art57}

\bibitem{curtarolo:art58}
W.~Setyawan and S.~Curtarolo, \emph{High-throughput electronic band structure
  calculations: Challenges and tools}, Comput.\ Mater.\ Sci. \textbf{49},
  299--312 (2010).
\input{curtarolo:art58}

\bibitem{curtarolo:art49}
O.~Levy, G.~L.~W. Hart, and S.~Curtarolo, \emph{Uncovering Compounds by Synergy
  of Cluster Expansion and High-Throughput Methods}, J.\ Amer.\ Chem.\ Soc.
  \textbf{132}, 4830--4833 (2010).
\input{curtarolo:art49}

\bibitem{curtarolo:art87}
G.~L.~W. Hart, S.~Curtarolo, T.~B. Massalski, and O.~Levy, \emph{Comprehensive
  Search for New Phases and Compounds in Binary Alloy Systems Based on
  {P}latinum-Group Metals, Using a Computational First-Principles Approach},
  Phys.\ Rev.\ X \textbf{3}, 041035 (2013).
\input{curtarolo:art87}

\bibitem{curtarolo:art127}
A.~R. Supka, T.~E. Lyons, L.~S.~I. Liyanage, P.~{D'{A}mico},
  R.~{Al~Rahal~Al~Orabi}, S.~Mahatara, P.~Gopal, C.~Toher, D.~Ceresoli,
  A.~Calzolari, S.~Curtarolo, M.~{Buongiorno Nardelli}, and M.~Fornari,
  \emph{{\small AFLOW}$\pi$: A minimalist approach to high-throughput {\it ab
  initio} calculations including the generation of tight-binding hamiltonians},
  Comput.\ Mater.\ Sci. \textbf{136}, 76--84 (2017).
\input{curtarolo:art127}

\bibitem{curtarolo:art121}
M.~J. Mehl, D.~Hicks, C.~Toher, O.~Levy, R.~M. Hanson, G.~L.~W. Hart, and
  S.~Curtarolo, \emph{The {AFLOW} Library of Crystallographic Prototypes: Part
  1}, Comput.\ Mater.\ Sci. \textbf{136}, S1--S828 (2017).
\input{curtarolo:art121}

\bibitem{curtarolo:art75}
S.~Curtarolo, W.~Setyawan, S.~Wang, J.~Xue, K.~Yang, R.~H. Taylor, L.~J.
  Nelson, G.~L.~W. Hart, S.~Sanvito, M.~{Buongiorno Nardelli}, N.~Mingo, and
  O.~Levy, \emph{{AFLOWLIB.ORG}: A distributed materials properties repository
  from high-throughput {\it ab initio} calculations}, Comput.\ Mater.\ Sci.
  \textbf{58}, 227--235 (2012).
\input{curtarolo:art75}

\bibitem{curtarolo:art92}
R.~H. Taylor, F.~Rose, C.~Toher, O.~Levy, K.~Yang, M.~{Buongiorno Nardelli},
  and S.~Curtarolo, \emph{A {REST}ful {API} for exchanging materials data in
  the {AFLOWLIB}.org consortium}, Comput.\ Mater.\ Sci. \textbf{93}, 178--192
  (2014).
\input{curtarolo:art92}

\bibitem{curtarolo:art128}
F.~Rose, C.~Toher, E.~Gossett, C.~Oses, M.~{Buongiorno Nardelli}, M.~Fornari,
  and S.~Curtarolo, \emph{{AFLUX}: The {LUX} materials search {API} for the
  {AFLOW} data repositories}, Comput.\ Mater.\ Sci. \textbf{137}, 362--370
  (2017).
\input{curtarolo:art128}

\bibitem{kresse_vasp}
G.~Kresse and J.~Hafner, \emph{{\it Ab initio} molecular dynamics for liquid
  metals}, Phys.\ Rev.\ B \textbf{47}, 558--561 (1993).
\input{kresse_vasp}

\bibitem{PAW}
P.~E. Bl\"{o}chl, \emph{Projector augmented-wave method}, Phys.\ Rev.\ B
  \textbf{50}, 17953--17979 (1994).
\input{PAW}

\bibitem{PBE}
J.~P. Perdew, K.~Burke, and M.~Ernzerhof, \emph{Generalized Gradient
  Approximation Made Simple}, Phys.\ Rev.\ Lett. \textbf{77}, 3865--3868
  (1996).
\input{PBE}

\bibitem{Jafarov_im_2014}
Y.~I. Jafarov, S.~Z. Imamalieva, V.~P. Zlomanov, and M.~B. Babanly, \emph{Phase
  equilibria in the reciprocal system
  3{Tl}$_2${S}+{Bi}$_2${Te}$_3$$\leftrightarrow$3{Tl}$_2$Te+{Bi}$_2$S$_3$},
  Inorg. \ Mater. \textbf{50}, 551--558 (2014).
\input{Jafarov_im_2014}

\bibitem{Babanly_amchemscij_2016}
M.~B. Babanly, Y.~I. Jafarov, Z.~S. Aliev, and I.~R. Amiraslanov,
  \emph{Chemistry of Thallium-based Topological Insulators}, ACSJ
  \textbf{10(1)}, 1--13 (2016).
\input{Babanly_amchemscij_2016}

\bibitem{cottrell_1967}
A.~H. Cottrell, \emph{An introduction to metallurgy} (St. Martin'{s} Press, UK,
  1967).
\input{cottrell_1967}

\bibitem{Burton_jap_2006}
B.~P. Burton, A.~{van de Walle}, and U.~Kattner, \emph{First-principles phase
  diagram calculations for the wurtzite-structure systems {Al}{N}-{Ga}{N},
  {Ga}{N}-{In}{N}, and {Al}{N}-{In}{N}}, J.\ Appl.\ Phys. \textbf{100}, 113528
  (2006).
\input{Burton_jap_2006}

\bibitem{Adjaoud_prb_2009}
O.~Adjaoud, G.~Steinle-Neumann, B.~P. Burton, and A.~{van de Walle},
  \emph{First-principles phase diagram calculations for the {Hf}{C}-{Ti}{C},
  {Zr}{C}-{Ti}{C}, and {Hf}{C}-{Zr}{C} solid solutions}, Phys.\ Rev.\ B
  \textbf{80}, 134112 (2009).
\input{Adjaoud_prb_2009}

\bibitem{Liu_chemgeo_2016}
Z.~T.~Y. Liu, B.~P. Burton, S.~V. Khare, and P.~Sarin, \emph{First-principles
  phase diagram calculations for the carbonate quasibinary systems
  {Ca}{C}{O}$_3$-{Zn}{C}{O}$_3$, Cd{C}{O}$_3$-{Zn}{C}{O}$_3$,
  {Ca}{C}{O}$_3$-Cd{C}{O}$_3$ and {Mg}{C}{O}$_3$-{Zn}{C}{O}$_3$}, Chem.\ Geol.
  \textbf{443}, 137--145 (2016).
\input{Liu_chemgeo_2016}

\bibitem{Kanai_JACS_Spinodal_2004}
T.~Kanai, T.~Sawada, J.~Yamanaka, and K.~Kitamura, \emph{Equilibrium
  Characteristic at Ordered-Disordered Phase Boundary in Centrifuged
  Nonequilibrium Colloidal-Crystal System}, J.\ Amer.\ Chem.\ Soc.
  \textbf{126}, 13210--13211 (2004).
\input{Kanai_JACS_Spinodal_2004}

\bibitem{deLaFiguera_PRL_Spinodal_2008}
J.~de~la Figuera, F.~L\'{e}onard, N.~C. Bartelt, R.~Stumpf, and K.~F. McCarty,
  \emph{Nanoscale Periodicity in Stripe-Forming Systems at High Temperature:
  Au/W$(110)$}, Phys.\ Rev.\ Lett. \textbf{100}, 186102 (2008).
\input{deLaFiguera_PRL_Spinodal_2008}

\bibitem{Chang_prb_2011}
J.~Chang, L.~F. Register, S.~K. Banerjee, and B.~Sahu, \emph{Density functional
  study of ternary topological insulator thin films}, Phys.\ Rev.\ B
  \textbf{83}, 235108 (2011).
\input{Chang_prb_2011}

\bibitem{Strained_Layer_Superlattices_32}
T.~P. Pearsall, ed., \emph{Strained-Layer Superlattices:Physics}, vol.~32
  (Academic Press, New York, 1990).
\input{Strained_Layer_Superlattices_32}

\bibitem{Strained_Layer_Superlattices_33}
T.~P. Pearsall, ed., \emph{Strained-Layer Superlattices:Materials Science and
  Technology}, vol.~33 (Academic Press, New York, 1991).
\input{Strained_Layer_Superlattices_33}

\bibitem{Kuroda10prl}
K.~Kuroda, M.~Ye, A.~Kimura, S.~V. Eremeev, E.~E. Krasovskii, E.~V. Chulkov,
  Y.~Ueda, K.~Miyamoto, T.~Okuda, K.~Shimada, H.~Namatame, and M.~Taniguchi,
  \emph{Experimental Realization of a Three-Dimensional Topological Insulator
  Phase in Ternary Chalcogenide {TlBiSe$_2$}}, Phys.\ Rev.\ Lett. \textbf{105},
  146801 (2010).
\input{Kuroda10prl}

\bibitem{Eremeev_jetpl_2010}
S.~V. Eremeev, Y.~M. Koroteev, and E.~V. Chulkov, \emph{Ternary Thallium-Based
  Semimetal Chalcogenides {Tl}-{V}-{VI}$_2$ as a New Class of Three-Dimensional
  Topological Insulators}, JETP\ Lett. \textbf{91}, 594 (2010).
\input{Eremeev_jetpl_2010}

\bibitem{Eremeev_prb_2011}
S.~V. Eremeev, G.~{Bi}hlmayer, M.~Vergniory, Y.~M. Koroteev, T.~V.
  Menshchikova, J.~Henk, A.~Ernst, and E.~V. Chulkov, \emph{Ab initio
  electronic structure of thallium-based topological insulators}, Phys.\ Rev.\
  B \textbf{83}, 205129 (2011).
\input{Eremeev_prb_2011}

\bibitem{Zhang_acsnano_2012}
Q.~Zhang, Z.~Zhang, Y.~Zhu, U.~Schwingenschl\"{o}gl, and Y.~Cui, \emph{Exotic
  Topological Insulator States and Topological Phase Transition in
  {Sb}$_2${Se}$_3$-{Bi}$_2${Se}$_3$ Heterostructures}, ACS\ Nano \textbf{6},
  2345 (2012).
\input{Zhang_acsnano_2012}

\bibitem{Jin_prb_2016}
K.-H. Jin, H.~W. Yeom, and S.-H. Jhi, \emph{Band structure engineering of
  topological insulator heterojunctions}, Phys.\ Rev.\ B \textbf{93}, 075308
  (2016).
\input{Jin_prb_2016}

\bibitem{Hutasoit_prb_2011}
J.~A. Hutasoit and T.~D. Stanescu, \emph{Induced spin texture in
  semiconductor/topological insulator heterostructures}, Phys.\ Rev.\ B
  \textbf{84}, 085103 (2011).
\input{Hutasoit_prb_2011}

\bibitem{Seixas_nc_2015}
L.~Seixas, D.~West, A.~Fazzio, and S.~B. Zhang, \emph{Vertical twinning of the
  Dirac cone at the interface between toplogical insulators and
  semiconductors}, Nat.\ Commun. \textbf{6}, 7630 (2015).
\input{Seixas_nc_2015}

\bibitem{Takahashi_prl_2011}
R.~Takahashi and S.~Murakami, \emph{Gapless Interface States between
  Topological Insulators with Opposite Dirac Velocities}, Phys.\ Rev.\ Lett.
  \textbf{107}, 166805 (2011).
\input{Takahashi_prl_2011}

\bibitem{Beule_prb_2013}
C.~{De~Beule} and B.~Partoens, \emph{Gapless interface states at the junction
  between two topological insulators}, Phys.\ Rev.\ B \textbf{87}, 115113
  (2013).
\input{Beule_prb_2013}

\bibitem{Johannsen_prb_2015}
J.~C. Johannsen, G.~{Au}t\'{e}s, A.~Crepaldi, S.~Moser, B.~Casarin, F.~Cilento,
  M.~Zacchigna, H.~Berger, A.~Magrez, P.~Bugnon, J.~Avila, M.~C. Asensio,
  F.~Parmigiani, O.~V. Yazyev, and M.~Grioni, \emph{Engineering the topological
  surface states in the ({Sb}$_2$)$_m$-{Sb}$_2${Te}$_3$ ($m$=0-3) superlattice
  series}, Phys.\ Rev.\ B \textbf{91}, 201101(R) (2015).
\input{Johannsen_prb_2015}

\bibitem{Gibson_prb_2013}
Q.~D. Gibson, L.~M. Schoop, A.~P. Weber, H.~Ji, S.~Nadj-Perge, I.~K. Drozdov,
  H.~Beidenkopf, J.~T. Sadowski, A.~Fedorov, A.~Yazdani, T.~Valla, and R.~J.
  Cava, \emph{Termination-dependent topological surface states of the natural
  superlattice phase {Bi}$_4${Se}$_3$}, Phys.\ Rev.\ B \textbf{88}, 081108(R)
  (2013).
\input{Gibson_prb_2013}

\end{thebibliography}

\newcommand{\Ozolins}{Ozoli\c{n}\v{s}}

\end{document}